\documentclass{kluwer}    
\usepackage{graphicx}
\newdisplay{guess}{Conjecture}
\begin{document}                               
\begin{article}
\begin{opening}         
\title{Dynamical Evolution of Barred Galaxies} 
\author{E. \surname{Athanassoula}}  
\runningauthor{E. Athanassoula}
\runningtitle{Dynamical Evolution of Barred Galaxies}
\institute{Observatoire de Marseille, 2 Place Le Verrier, 13248
  Marseille C\'edex 04, France}
\date{August, 2004}

\begin{abstract}
Angular momentum redistribution within barred galaxies drives their 
dynamical evolution. Angular momentum is emitted mainly by
near-resonant material in the bar region and absorbed by resonant
material mainly in the outer disc and in the halo. This exchange determines
the strength of the bar, the decrease of its pattern speed, as well as
its morphology. If the galaxy has also a gaseous component and/or a
companion or satellite, then these also take part in the angular
momentum exchange. During the evolution a bar structure forms in the
inner parts of the halo as well. This bar is shorter and 
fatter than the disc bar and stays so all through the simulation, 
although its length grows considerably with time. Viewed edge-on, 
the bar in the disc component acquires a boxy or peanut shape.
I describe the families of periodic orbits that 
explain such structures and review the observations showing that 
boxy/peanut `bulges' are in fact just bars seen edge-on. 
\end{abstract}
\keywords{barred galaxies, dynamical evolution, halo, peanuts, bulges}

\end{opening}           

\section{Introduction} 

Bars are very common features in disc galaxies. \citeauthor{Eskr+}
\shortcite{Eskr+}, using 
a statistically well-defined sample of 186 disc galaxies from the Ohio
State University Bright Spiral Galaxy Survey, find that 56\% are
strongly barred in the H band, while another 16\% are weakly
barred. \citeauthor{GrPP} \shortcite{GrPP}, using a smaller sample of
53 spirals observed in 
the K band, find that about 75\% of them have bars or ovals. These
bars can have very different morphologies, from short to long, and
from fat to thin, or from rectangular-like to elliptical-like. Thus
we need to explain not only their formation, but also the distribution
of their parameters and their different morphologies.
 
The spontaneous formation of bars in galactic discs was already shown
in $N$-body simulations of the early seventies (e.g. \citeauthor{mpq70}, 
\citeauthor{Hohl} \citeyear{Hohl}). 
At that time, the observational evidence for the 
existence of dark haloes around individual galaxies was
hardly compelling, so the discs in these simulations were
self-gravitating. Only a few years later haloes were propelled
into the center of scientific discussions. \citeauthor{OstrP} 
\shortcite{OstrP} were the first to check the effect of a heavy
halo on the bar instability and found it to be stabilising. Although
the number of particles in their simulations did not exceed 500, their
work is very insightful. They introduced the parameter $t_{op}$, which
is the ratio of kinetic energy of rotation to total gravitational
energy. They concluded that halo-to-disc mass ratios of 1 to 2.5 and an
initial value of $t_{op} = 0.14 \pm 0.03$ are required for
stability. Several later papers (e.g. \citeauthor{AS} \citeyear{AS})
confirmed the stabilising tendency of the halo. Yet, as we will see in the
next section, this is an artifact, due to the fact that these
simulations were either 2D, or had a rigid halo, or had too few
particles. Thus the halo was not properly described. Before I 
discuss the effect of a properly described halo, let me first
describe briefly a temporal 
sequence in the bar formation and evolution process.

Most $N$-body simulations start with a thin axisymmetric disc immersed
in a more or less massive halo. The initial density and velocity
distributions are chosen so as to be in agreement with observations of nearby
spirals. Of course such initial conditions are idealised and may not hold
for real galaxies, where the bar could well form at the same time as the
disc. Yet, following the disc and bar formation ab initio, in a
cosmological scenario, is a still largely unsolved problem, since it
requires knowledge of the initial conditions of the mass distribution
right after the formation of a disc galaxy. 
Furthermore, using present day initial conditions makes it possible
to study best the angular momentum exchange and to make comparisons with
analytical calculations. Nevertheless, problems like bar formation in
a growing disc, non-spherical and/or clumpy haloes and the effect of
infalling small satellites and debris need to be soon addressed.    

\begin{figure}
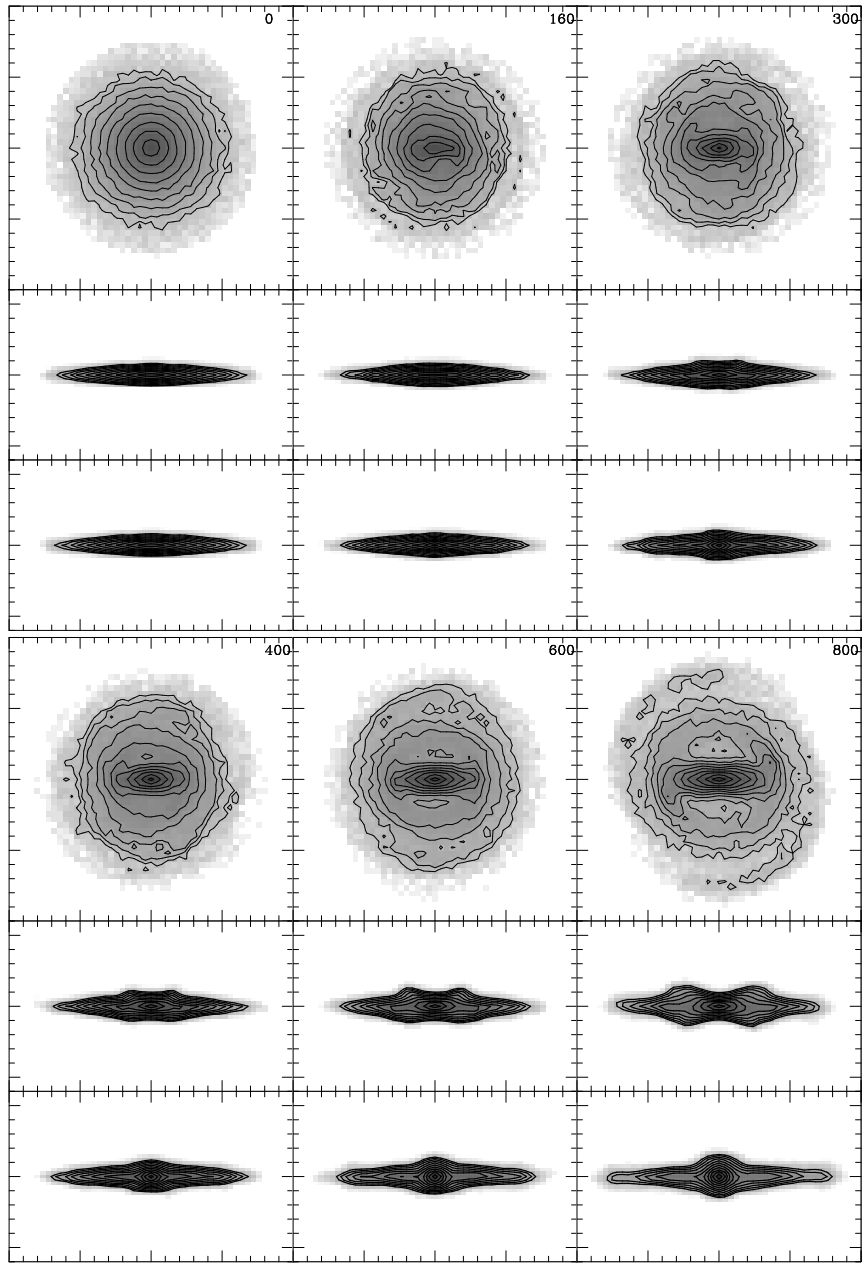
 
\rotatebox{0}{\includegraphics[width=27pc]{fig1a.ps}}
\rotatebox{0}{\includegraphics[width=27pc]{fig1b.ps}}
\caption[]{Formation of a bar in an initially axisymmetric disc. The
  upper and fourth rows give 
  the face-on views; the second and fifth ones the side-on views and
  the third and sixth rows the end-on views. Time
  increases from left to right and from top to bottom and is given in
  the upper right corner of each face-on panel. Here and elsewhere in 
  this paper, times are given in computer units as defined by 
  \inlinecite{AM02}.}
\label{fig:evol}
\end{figure}

Bar formation in such idealised initial conditions is illustrated in
Fig.~\ref{fig:evol}, which gives the face-on, the side-on and the
edge-on views of the disc component
at various times. I call side-on the edge-on view where
the line-of-sight is along the bar minor axis, while the end-on view is
also edge-on but with the line-of-sight along the bar major axis. At
the beginning the disc starts axisymmetric, but soon forms spirals
(not shown here), whose multiplicity depends on the disc-to-halo mass
ratio within the outer disc radius. 
The actual bar formation time also depends on the relative
importance of the halo component, being shorter in simulations whose
inner parts are more disc dominated. In the second of the times
illustrated the bar is still short and rather asymmetric. With time,
it grows more symmetric, but also longer, i.e. stronger. At the last
time shown, a ring structure encircles the bar. It is somewhat
elongated, but not far from circular, as inner rings observed in
barred galaxies \cite{Buta}. In all simulations the bar is direct,
i.e. rotates in the same sense as the particles in the disc, and
stops within corotation, in good agreement with periodic orbit
calculations \cite{C80}. 

Viewed edge-on, the disc shows also clear evolution. During the first
stages of its formation, the bar stays thin, but thickens shortly
thereafter. Its vertical extent increases as the bar strength
increases. Initially its shape is boxy and then gradually evolves to
peanut-like. Seen end-on the bar looks round, reminiscent of a bulge
in the central part of a disc. Seen this way also, the bar size
increases as it strengthens in the face-on view.

\section{Angular momentum exchange}

In a ground-braking paper, \citeauthor{LBK} \shortcite{LBK} discuss the
transport of angular momentum within discs of spiral galaxies. Using a
full analytical treatment, as well as a simple but insightful
individual orbit approach, they show that disc stars at inner Lindblad
resonance (hereafter ILR) emit angular momentum, while disc stars at
corotation (hereafter CR) and outer Lindblad resonance (hereafter OLR)
absorb it, so that angular momentum is transported outwards. If
the spiral amplitude is growing, material which is not at
near-resonance can also emit, or absorb, angular momentum.  

\citeauthor{Ath03} \shortcite{Ath03} extended the work 
of \citeauthor{LBK} to
include a spheroidal component. If the distribution function of this
component depends only on the energy, then all halo resonances will
absorb angular momentum. Thus the picture of \citeauthor{LBK} can be
generalised as follows : Angular momentum is emitted from material at
near-resonance in the bar
region (ILR and other such resonances within CR, e.g. 1:4; see Appendix
for the notation) and absorbed by material at resonance in the outer
disc and in the halo. Colder material can emit/absorb more angular
momentum that hotter one. Thus the halo is less responsive than the disc
per equal amount of resonant mass. However, there is not much material
in the outer disc, where the density is very low, while the halo can
be very massive. It can thus be that the halo absorbs more angular
momentum than the outer disc, and this has proven to be the case in
many $N$-body simulations, as will be discussed in the next
section. Since the bar is 
a negative angular momentum `perturbation' (\citeauthor{Kalnajs}
\citeyear{Kalnajs}, \citeauthor{LBK} \citeyear{LBK}), by losing angular
momentum it becomes stronger. 

\citeauthor{TW84} \shortcite{TW84} 
investigated dynamical friction on a solid bar
rotating in a spherical halo and found that this arises from halo
stars which are near-resonant with the rotating bar. They derived an
analogue of Chandrasekhar's formula \cite{Chandra} for spherical systems,
valid when the angular velocity of the bar does not change too
slowly. \citeauthor{W85} \shortcite{W85} calculated that this will
cause a considerable slow-down of the bar within a few bar
rotations. As we will see in the next section, $N$-body simulations also
present such a slow down (e.g. \citeauthor{LC1} \citeyear{LC1},
\citeauthor{LC2} \citeyear{LC2}, \citeauthor{HW92} \citeyear{HW92},
\citeauthor{Ath96} \citeyear{Ath96}, \citeauthor{DebSel} \citeyear{DebSel},
\citeauthor{VK} \citeyear{VK}),
but considerably less than that predicted by
\citeauthor{W85} \shortcite{W85}. \citeauthor{W04} \shortcite{W04}
stressed that the derivation of the formulae of \citeauthor{LBK}
assumes that the perturbations grow slowly over a very long time and 
that transients can be ignored. Their quantitative results should 
thus not be relied upon, since these assumptions may not hold in
real galaxies. This, however, does not affect any of the results
reviewed in this section, since they do not rely on any quantitative
use of the \citeauthor{LBK} formalism.

\citeauthor{Fuchs} \shortcite{Fuchs} showed that the dynamics of a
self-gravitating shearing sheet are strongly modified if this is
immersed in a live halo. Namely, the amplitude of the density wave is
very considerably enhanced.This work is presently being generalised to
non-isotropic velocity distributions.

\section{Results from $N$-body simulations}

\begin{figure} 
\rotatebox{-90}{\includegraphics[width=20pc]{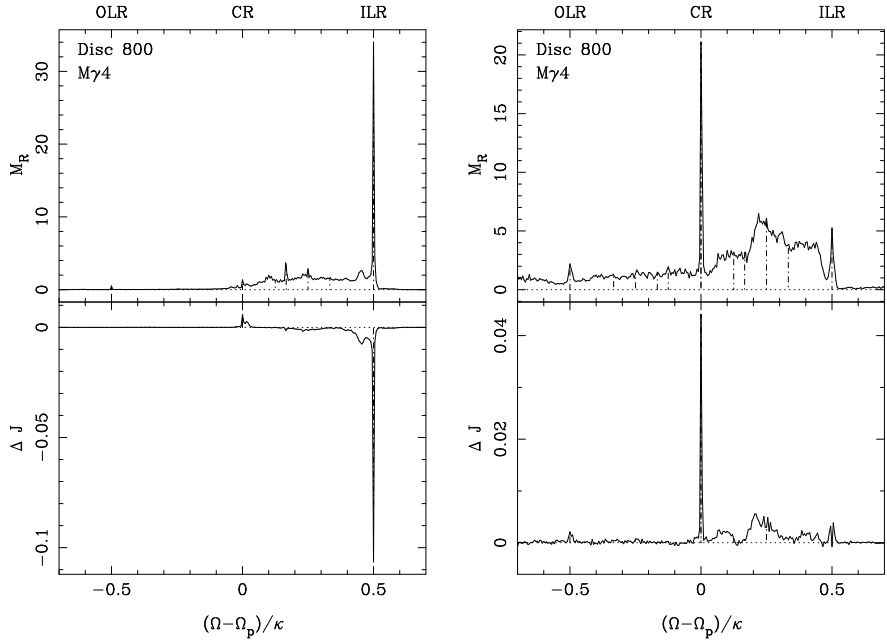}}
\caption[]{Resonances in the disc and the spheroidal component. The
  upper panels give, for the time $t$ = 800, the mass per unit
  frequency ratio, $M_R$, as a function of that ratio. The frequency
  ratio is defined as $(\Omega - \Omega_p) / \kappa $, the ratio of the
  angular frequency in the frame in which the bar is at rest to the
  epicyclic frequency. The lower
  panels give $\Delta J$, the angular momentum gained or lost by
  particles of a given frequency ratio between times 500 and 800, as
  a function of that frequency ratio, calculated at $t$ = 800. The
  left panels correspond to the disc component and the right ones to
  to the spheroid. The vertical dot-dashed lines give the positions
  of the main resonances. The component and time are given in 
  each panel.This figure is similar to figure 1 of
  \inlinecite{Ath03},
  but now more particles have been used in the frequency
  calculations of the spheroidal component, so as to improve the signal
  to noise ratio. This is particularly clear in the lower right panel.
}
\label{fig:resonances}
\end{figure}
Contrary to real galaxies, $N$-body simulations are well suited for
studying the angular momentum exchange within a galaxy. This has been
one of the main goals of \inlinecite{Ath02} and \inlinecite{Ath03} 
and I will retrace here a number
of the steps made in those papers.

The first point to check is that there is a considerable amount of
near-resonant material in the halo component. This is shown in the
upper panels of
Fig.~\ref{fig:resonances}, where I plot the mass per unit frequency
ratio, $M_R$, as a function of the frequency ratio $(\Omega -
\Omega_p) / \kappa$ (see appendix for the notation). It is clear that
the distribution is not uniform, and has peaks at the resonances. In
all simulations, the disc has a strong peak at ILR, which is made of
particles trapped around this resonance and constituting the backbone
of the bar. Secondary peaks can be found
at other resonances, as e.g. inner 1:3, inner 1:4, CR or OLR, whose
existence and height varies from one simulation to another. More
important, the halo component also shows similar peaks. The highest is
at CR, while secondary peaks can be seen at ILR and OLR. Such peaks
can be seen in all simulations I analysed, again with varying
heights. 

The bottom panels of Fig.~\ref{fig:resonances} show the way the
angular momentum is exchanged. For the disc component it is emitted
from the region within the bar, and particularly the ILR, and absorbed
at CR (and in some simulations also at OLR). However, the amount of
angular momentum emitted is much more than what the outer disc
absorbs. This is understood with the help of the bottom right panel,
which shows that all the halo resonances absorb a considerable amount
of angular momentum, much more so than the outer disc. Thus simulations
confirm the angular momentum exchange mechanism suggested by the
analytical work, and show that the halo can be an important agent
in this respect.

\begin{figure} 
\rotatebox{-90}{\includegraphics[width=20pc]{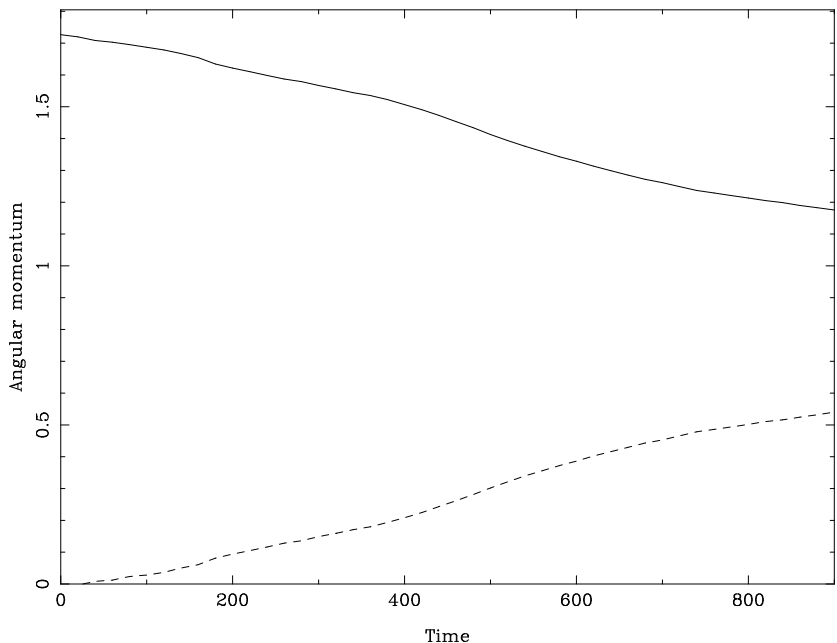}}
\caption[]{Angular momentum of the disc (full line) and spheroidal
  component (dashed line) as a function of time for a simulation 
  producing a strong bar component.}
\label{fig:tangmom}
\end{figure}

The global redistribution of angular momentum is seen in
Fig.~\ref{fig:tangmom}. This shows that the disc component loses a
considerable amount of angular momentum which is taken by the
halo. The amount of course varies from one simulation to another and
depends on the parameters of the disc and the halo. Such a
redistribution was also seen by \inlinecite{DebSel} and
by \inlinecite{VK}.

\begin{figure} 
\rotatebox{0}{\includegraphics[width=27pc]{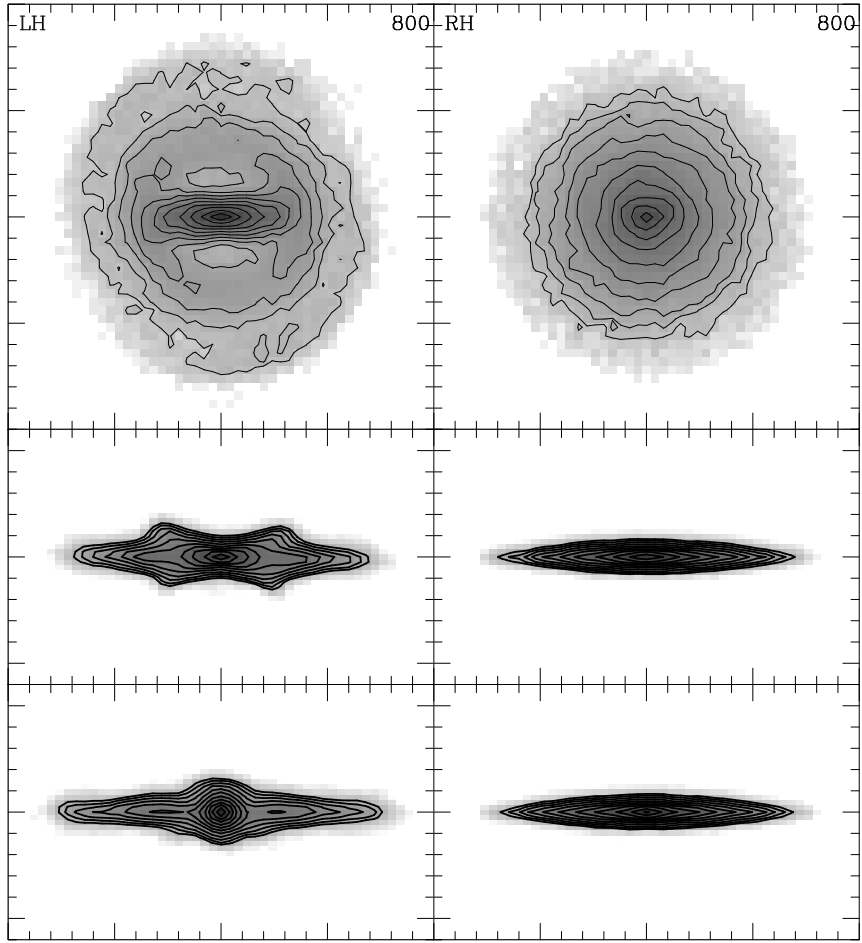}}
\caption[]{Comparison of two simulations with initially identical disc
  components and haloes with identical mass distributions. The time 
  displayed is towards the end of the simulation. In the
  simulation displayed in the left
  panels the halo is live, i.e. it can respond to the disc evolution and
  exchange angular momentum with it. In the simulation in the right
  panels the halo is 
  rigid, i.e. it is just a superposed constant potential, so that no
  angular momentum exchange between the disc and the halo is
  possible. The time is given in computer units in the upper right
  corner of the face-on views}
\label{fig:LHRH}
\end{figure}

Fig.~\ref{fig:LHRH} shows that the angular momentum redistribution can
be very important for the bar evolution. It compares two simulations
with initially identical disc components and with haloes having 
initially identical mass distributions. In the simulation in the left
panels the halo is live (LH), i.e. it is composed of 
particles. In the simulation in the right panels, however, the halo is
rigid (RH), i.e. it is an imposed potential. Thus in LH the halo can absorb
angular momentum, while in RH it can not. The difference in the evolution
is very striking. Simulation LH grows a very strong bar, which, when
seen side-on, has a strong peanut shape. Simulation RH, on the contrary,
has a very mild oval in the innermost regions, and hardly evolves when
seen edge-on. The large difference between the results of the two simulations
argues strongly that the angular momentum absorbed by the halo
can be a decisive factor in the evolution of the bar component.

Theoretical arguments predict that the angular momentum emitted or
absorbed at a given resonance depends not only on the density of
matter there, but also on how cold the near-resonant material is. 
This is borne out by the simulations. Indeed, if the disc is hot
(i.e. has a high initial $Q$) and/or the halo is very hot, then the
bar does not grow to be very strong and does not slow down much.
Examples of this are given by \inlinecite{Ath03}.

\begin{figure} 
\rotatebox{0}{\includegraphics[width=27pc]{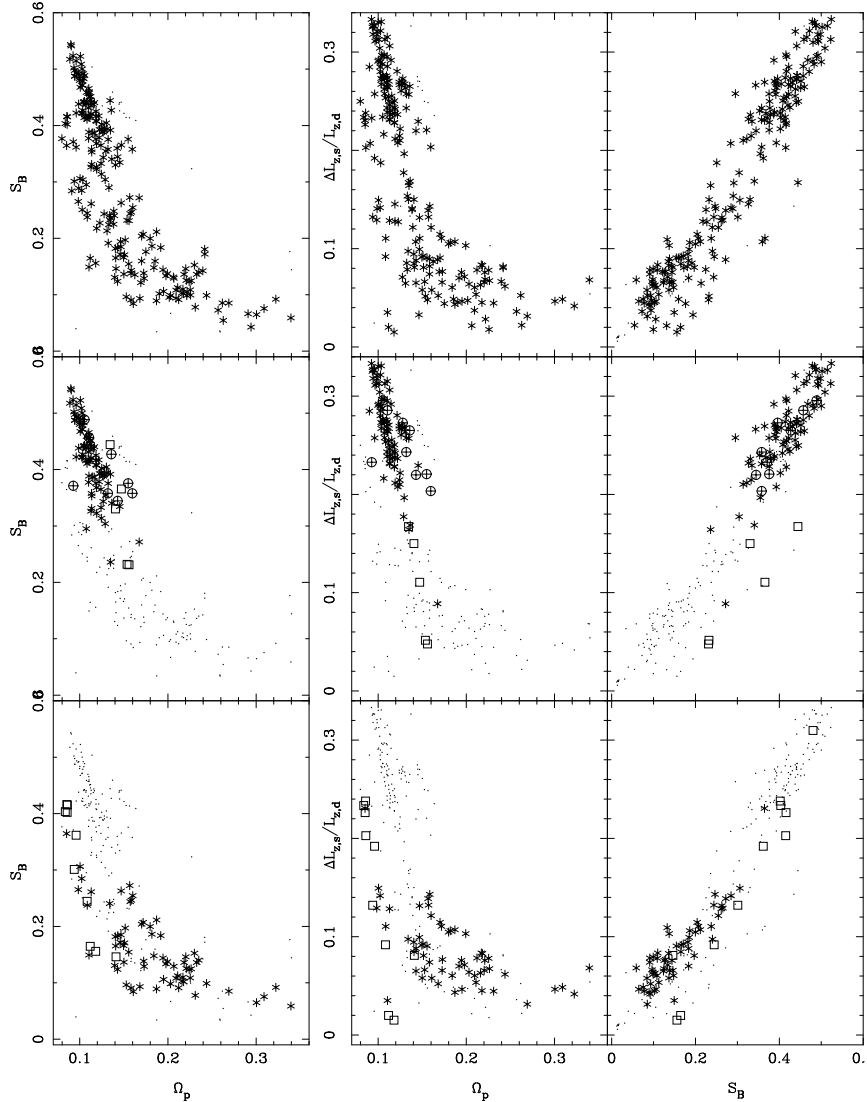}}
\caption[]{Relations between the bar strength and its pattern speed
  (left panels), the angular momentum acquired by the spheroid and
  the bar pattern speed (middle panels) and the angular
  momentum of the spheroid and the bar strength (right panels), at
  time $t$ = 800. The
  spheroid angular momentum is normalised to the initial disc angular
  momentum ($L_{z,d}$). Each symbol represents the results of a
  simulation. The upper row includes all simulations. In the middle
  one, simulations with a concentrated halo (MH types in the notation
  of \citeauthor{AM02} \citeyear{AM02}) are shown with a large symbol
  and those with a non-concentrated halo (MD types in the notation
  of \citeauthor{AM02} \citeyear{AM02}) with a
  dot. MH simulations with a bulge are plotted with a circled cross,  
  those with an initially very hot disc with an open square and the
  rest with a star. In the lower row simulations with a
  non-concentrated halo (MD types) are shown with a large symbol  
  and those with a concentrated halo (MH types) with a
  dot. MD simulations with an initially very hot disc are singled
  out with an open square and the rest are given with a star. }
\label{fig:correlations}
\end{figure}

In fact, it is the angular momentum redistribution that determines the
strength and the slowdown rate of the bar. This can be seen in
Fig.~\ref{fig:correlations}, which includes results (i.e. strength and
pattern speed of the bar and angular momentum absorbed by the halo) 
from a very large number of simulations, all with the same disc and 
halo masses and the same initial 
disc scale-length. The upper right panel shows that there is a
clear correlation between the strength of the bar and the angular
momentum absorbed by the halo component. The upper middle panel shows
the corresponding relation, now for the bar pattern speed. Let me now split
the sample in two, one including simulations with haloes which have
considerable mass in the regions of the main resonances, and one in
which the halo mass is mainly in the parts outside the main
resonances. In simulations in the 
first sample (middle row of panels), it is the halo
that absorbs most of the angular momentum emitted by the bar region. A
correlation between the pattern speed of the bar and the amount of
angular momentum in the halo is thus expected, and is indeed found. In 
simulations in the
second sample the disc plays a more important role in the angular
momentum exchange, and only a weak trend with the angular
momentum absorbed by the halo is expected, and this is indeed borne
out by the simulations.

Such correlations are expected for all types of disc and spheroid models.
The location of the regression line and of the individual points on the 
relevant planes -- e.g. the ($\Delta$L$_{z,s}$/L$_{z,d}$, S$_B$) plane,
where $\Delta$L$_{z,s}$ the amount of angular momentum absorbed by the
spheroid, L$_{z,d}$ the initial disc angular momentum, and S$_B$ the
bar strength -- may well change with the model used.

\begin{figure} 
\rotatebox{0}{\includegraphics[width=29pc]{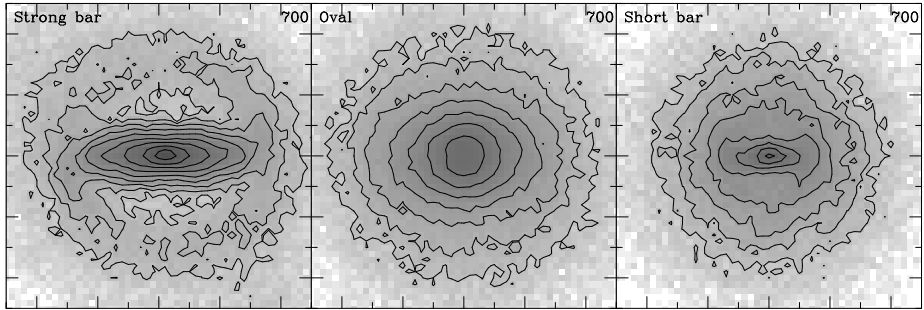}}
\caption[]{ Three distinct bar morphologies. In the left panel the bar
is strong, it has rectangular-like isodensities and is surrounded by 
a ring. In the
middle panel the bar is an oval, while in the right panel it is short.}
\label{fig:shapes}
\end{figure}

The angular momentum exchange also determines the morphology of the
bar. In all simulations where a considerable amount of angular
momentum is redistributed, the bar is strong, as the example
given in the left panel of Fig.~\ref{fig:shapes}. Such bars are long
and have rectangular-like isodensities (\citeauthor{AM02},
\citeyear{AM02}). They are often surrounded by a ring, whose location 
and shape is like those of the inner
ring in real galaxies. The properties of such simulated bars are
similar to those of strong bars in early type disc galaxies
\cite{Atokyo}. This can be easily explained, since such galaxies have big
bulges. These components are spheroids similar to the halo, and seen
their extent, they will have near-resonant material which will absorb
angular momentum, thus helping the bar become yet stronger than it would
in their absence.

Simulations in which little angular momentum has been exchanged do 
not form strong bars. Schematically, their morphology can be either
that of an oval (middle panel of Fig.~\ref{fig:shapes}), or that of a
short bar (right panel). Ovals are found predominantly in simulations
with hot discs. Since in such cases it is the halo that absorbs most
of the angular momentum, the bar can extend relatively far out 
in the disc to maximise the extent and therefore the contribution of
the emitting region. Short bars are found predominantly in
simulations with very hot, non-responsive haloes. Then the angular
momentum absorbed by the outer disc is a considerable fraction of the
total amount exchanged, and thus bars have to stay short in order to
leave sufficient mass in the outer disc regions to do the absorption. 
Thus simulations can
reproduce well the morphologies of observed bars and the wide range of
bar strengths observed. 

In  
\inlinecite{Ath03} only the disc and halo (and sometimes bulge)
components are taken into account. Yet the complete picture of angular momentum
exchange can be more complicated. Galaxies (particularly late types)
have also a gaseous disc component. This may give angular momentum to
the bar, and thus decrease its strength \cite{BAHF}. Furthermore,
galaxies are not isolated universes, and thus can interact with their
companions, or satellites. If the latter absorb angular momentum, then the
bar can grow stronger than in the isolated disc \cite{BAHF}. This is
in good agreement with observational results that show that more bars
can be found in interacting than in isolated galaxies \cite{Elme}.
 
\section{A bar in the halo component}

\begin{figure}
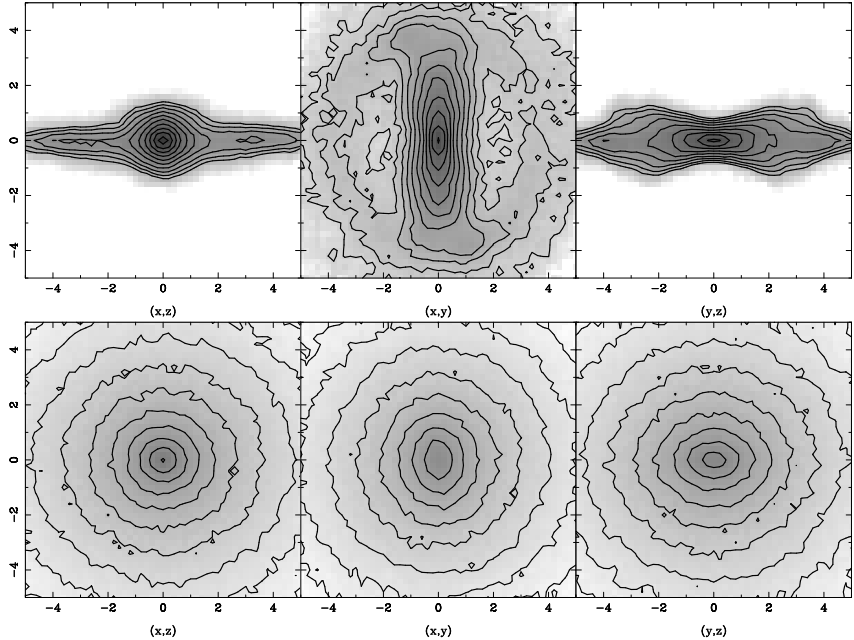
 
\rotatebox{-90}{\includegraphics[width=10pc]{fig7a.ps}}
\rotatebox{-90}{\includegraphics[width=10pc]{fig7b.ps}}
\caption[]{ Three orthogonal views of the disc (upper panels) and halo
components (lower panels). The central panel is a face-on view, while
the two others are edge-on; side-on for the right panels and end-on for
the left ones.Note that the halo component does not stay axisymmetric, 
but forms an oval in its inner parts, to which I refer to here as the
halo bar.}
\label{fig:halo-disc}
\end{figure}

Since the halo is not rigid, it evolves dynamically together with
the disc. In simulations in which a considerable amount of angular
momentum has been exchanged and which have thus formed a strong bar,
the halo does not 
stay axisymmetric. It also forms a bar, or more precisely an oval,
which I will call, for brevity, the halo bar. An example is seen in
Fig.~\ref{fig:halo-disc}, which compares the morphology of the disc
bar (upper panels) to that of the 
halo bar (lower panels). A very clear case of
such a structure is also seen in 
Fig. 2 of \inlinecite{HBKW}.

\begin{figure} 
\rotatebox{-90}{\includegraphics[width=20pc]{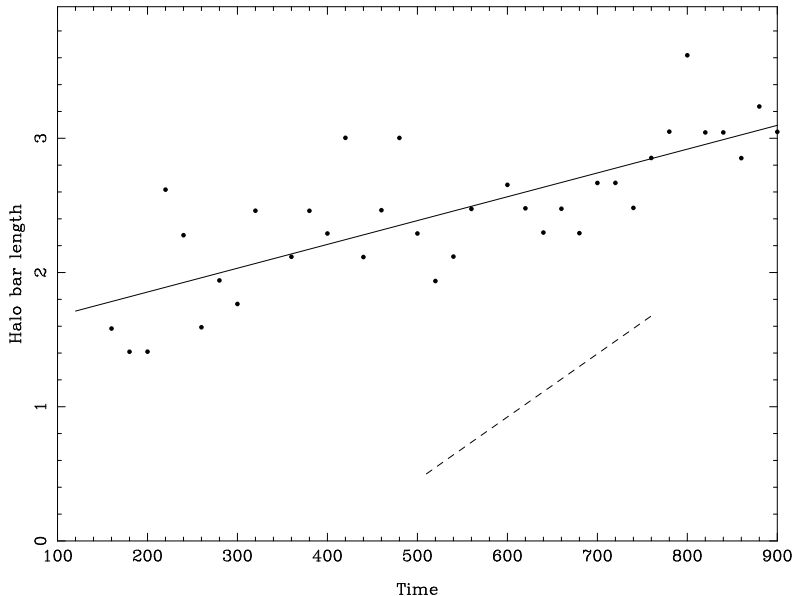}}
\caption[]{ Length of the halo bar as a function of time. The solid
  line is a least square fit. The slope of the dashed line gives the growth of
  the length of the disc bar with time.}
\label{fig:hbar-time}
\end{figure}

Halo bars are
triaxial, but nearer to prolate than to oblate, with their minor axis
perpendicular to the disc equatorial plane. The axial ratio in this plane
(ratio of minor to major axis) increases with increasing radius, so
that halo bars tend to become axisymmetric in the outer parts. They are always
considerably fatter than the corresponding disc bar, and thus nearer to
ovals. Since the change in axial ratio
is very gradual, it is not easy to define precisely the end of the
halo bar, and thus to calculate its length. 
It is clear, however, that it is always considerably
shorter than the disc bar. The length of the halo bar as a function of
time for a simulation with a strong disc bar
is given in Fig.~\ref{fig:hbar-time}. The
spread in the measurements reflects the difficulty of defining
precisely the end of the halo bar. Yet a least square fit shows
clearly that the length of the halo bar increases with time. The
slope of the dashed line gives the rate of increase of the disc bar length, and
thus shows that the length of the halo bar increases less fast than
that of the disc bar.  
The halo bar has roughly the same orientation as the disc bar at all
times. In good agreement with \inlinecite{HBKW}, I 
find that the m=2
component of the halo continues 
also outside the halo bar, trailing behind the disc bar.
More information on the halo bar properties, their relation with the
disc bar properties, and the changes in the halo orbital structure 
as a result of the halo bar formation will be given elsewhere 
\cite{Ath04b}.

\section{The formation of boxy/peanut features}

$N$-body simulations show that bars form initially in the disc 
equatorial plane. Very shortly thereafter, however, they buckle
out of this plane, creating initially an asymmetric structure
which soon evolves into a symmetric peanut 
(\citeauthor{ComSan} \citeyear{ComSan}, \citeauthor{CDFP} \citeyear{CDFP},
\citeauthor{RSJK} \citeyear{RSJK}, \citeauthor{AM02} \citeyear{AM02},
\citeauthor{Ath03} \citeyear{Ath03}, \citeauthor{OND} \citeyear{OND}, 
\citeauthor{Ath04a} \citeyear{Ath04a})
The formation of such a structure is easily understood in the framework
of orbital structure in 3D barred galaxy potentials. Such a study
was initiated by \inlinecite{P84}, who showed that there are
several families of 3D orbits, bifurcated at the vertical resonances
of the main planar family. This work was supplemented and extended
in a series of four papers (\citeauthor{SPA02a} \citeyear{SPA02a},
\citeyear{SPA02b}, \citeauthor{PSA02} \citeyear{PSA02},
\citeyear{PSA03}, also \opencite{Puebla}).
I will briefly recall here the main results of these papers that
are relevant to the formation of the peanut structure. I will
also follow their notation.

2D orbital structure studies clearly established that it is the
stable members of the x$_1$ family of periodic orbits (i.e. the
orbits that are elongated along the bar and close after one rotation
and two radial oscillations) that constitute the backbone of the bar
(\citeauthor{CP80} \citeyear{CP80}, \citeauthor{ABMP} \citeyear{ABMP}, 
etc.). Such periodic
orbits trap around them regular, non-periodic orbits and, due to their
appropriate size, shape and orientation, form the bar.

3D orbital studies, however, showed that the situation is much more
complicated. The backbone of 3D bars is the x$_1$ tree, i.e. the x$_1$
family plus a tree of 2D and 3D families bifurcating from it 
(\citeauthor{SPA02a} \citeyear{SPA02a},
\citeauthor{SPA02b} \citeyear{SPA02b}). 
Trapping around these families will determine the extent and 
shape of the bar in the face-on view, as well as the edge-on extent
and shape. However, it need not be the same family that determines
both. The radial extent and outer outline in the face-on view is 
usually determined by the x$_1$ family and/or 2D rectangular-like
orbits at the planar 4:1 resonance region. On the other hand, the
length of the peanut is determined by the extent of the 3D family
that constitutes its backbone, e.g. the x1v1, or the x1v4 family (see
\citeauthor{SPA02a} \citeyear{SPA02a} for more information on these
families and their properties). 
This means that the length of the peanut must be somewhat shorter 
than that of the bar seen face-on, the exact ratio depending on which
family(ies) contitute(s) the backbone of the peanut. It is important
to note that the (x,y) projection of the members of the families
of the x$_1$ tree has, in general, an outline which is very similar to
that of the
members of the x$_1$ family (2D orbits) at the same energy. This
means that the difference between the two stands out in the
edge-on view, but can be seen only little in the face-on view.

The vertical family which bifurcates from the x$_1$ at the lowest 
energy -- x1v1 in the notation of \inlinecite{SPA02a} -- is 
particularly useful
for building boxy or peanut-like features if it does not have a
sizeable complex unstable part. If this family is the backbone
of the peanut, then this feature is considerably shorter than 
the bar, the ratio of the two being less than 0.5 in the models
of \inlinecite{PSA02}. On the other hand, if the backbone is a family which
bifurcates at higher energies, then the peanut/bar ratio is 
considerably larger. E.g. if the backbone is the x1v4 family, then
the ratio is of order of 0.8. In this case the vertical extension
of the peanut or boxy feature is also considerably less. If more than
one vertical family contributes significantly to the peanut, it is
generally the one which bifurcates at the lowest energy that plays
the main role morphologically. Finally, it is worth mentioning that 3D
families which bifurcate from the x$_2$ family (i.e. from the
planar family which is elongated perpendicular to the bar) can also support
boxy morphologies. This can influence the morphology of bars 
seen end-on.

Such boxy or peanut structures, when seen in real galaxies, are
called by observers ``boxy or peanut bulges''. This is due to
the fact that, according to the simple morphological definition,
a bulge is a smooth light distribution that swells out of the
central part of the disc seen edge-on; a definition that is well
fulfilled by boxy/peanut structures. However, bulges are not a
homogeneous class of objects. As discussed by \inlinecite{Ath04a}, 
they include classical bulges, boxy/peanut features, and disc-like 
bulges. Classical bulges have spheroidal shapes, a maximum 
thickness (seen edge-on) at the center of the galaxy, and a 
projected light, or surface density, profile close
to the r$^{1\over4}$ law. They presumably form from hierarchical
merging and/or collapse. On the other hand, boxy/peanut features
are just part of the bar. In fact, as already discussed, they are
the part that is
composed of the vertically extending orbits from 3D families which
bifurcate from the x$_1$ family. Finally, disc-like bulges are
central disc-like objects formed from gaseous material pushed
inwards by torques due to bars, or perhaps other non-axisymmetric
features, which later formed stars. This last class of objects
has been mainly defined with the help of radial photometric
profiles, i.e. they are excess light in the central part of
disc galaxies, above the simple exponential profile fitting
the remaining, non-central part of the disc.

\inlinecite{Ath04a} presented a number of comparisons of 
observed properties of boxy/peanut features with those of edge-on 
bars and found excellent agreement. I will here discuss two such 
comparisons, namely the comparison of the features in median 
filtered images, and the features in the emission line and 
absorption line position -- velocity  diagrams,
and refer the reader to \inlinecite{Ath04a} for others.

\begin{figure} 
\rotatebox{0}{\includegraphics[width=28pc]{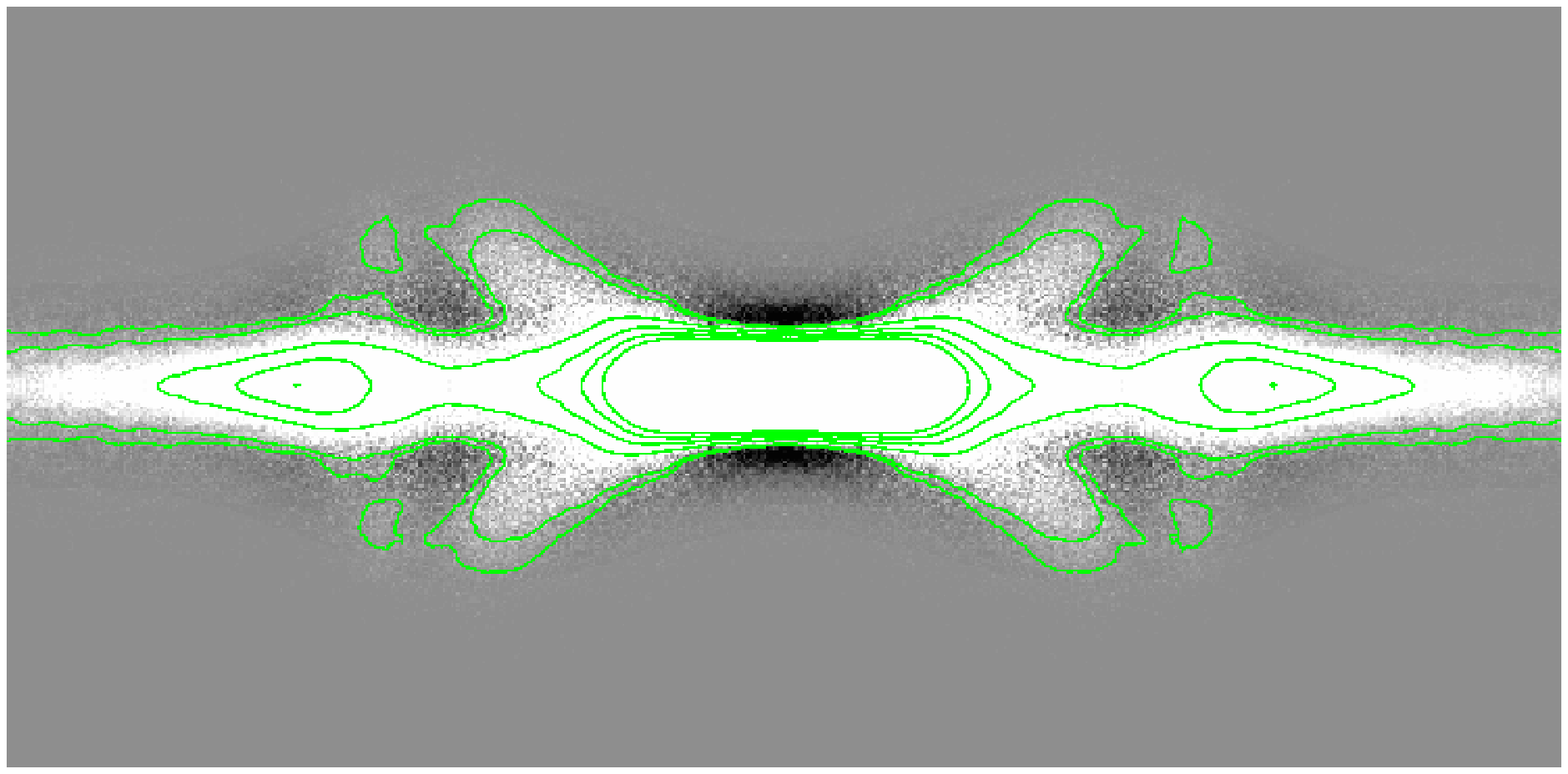}}
\rotatebox{0}{\includegraphics[width=28pc]{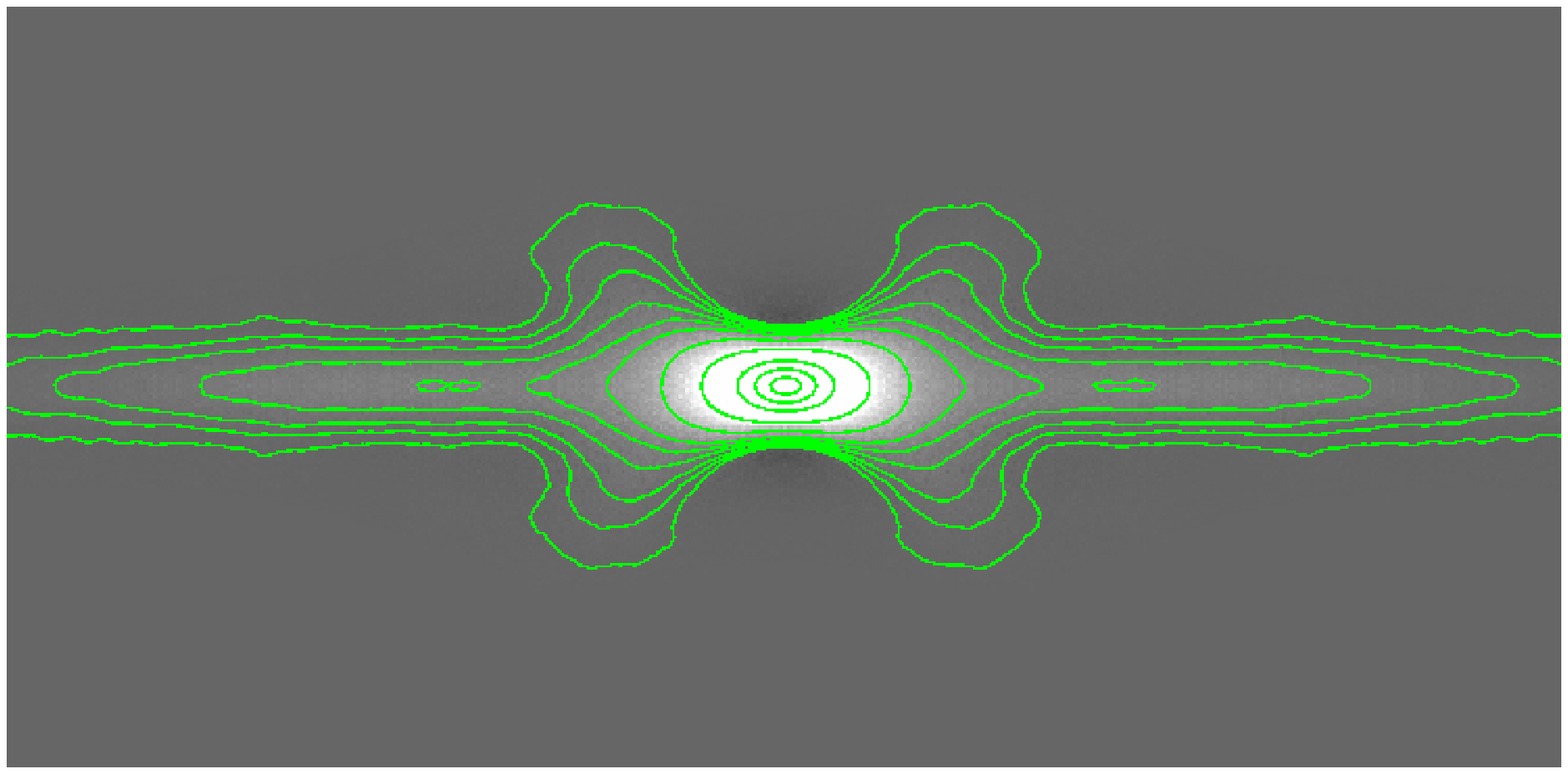}}
\caption[]{Median filtered image for the disc of a simulation with a
  strong bar. 
In the upper panel the bar is seen side-on, and in the lower panel
it is seen edge-on with the line-of-sight at 45$^\circ$ to the bar
major axis. The isodensities are chosen so as to show best the
relevant features.}
\label{fig:unsharp}
\end{figure}

\subsection{Median filtering}

A comparison of images of real galaxies and of simulated ones after comparable
median filtering is much more demanding than a comparison of 
the corresponding global
morphologies. This is due to the fact that median filtering reveals a
number of features, whose position and form have to be matched at a
given scale.

\citeauthor{Aron03} (\citeyear{Aron03}, \citeyear{Aron04}) have 
performed median filtering of near-infrared images of a sample of
edge-on boxy and peanut galaxies and found a number of interesting
features. In many cases (e.g. Fig. 2a in \opencite{Aron03}) 
one can note four extensions out of the 
equatorial plane, which form an X-like shape, except that the four
extensions do not necessarily cross the center. Another common feature
is maxima of the density along the equatorial plane, away from the
center and diametrically opposite. Namely, starting from the center of 
the galaxy and going outwards along the equatorial plane, the projected surface
density first drops and then increases again to reach a local maximum. It
then decreases to the edge of the disc. 

In order to compare the simulations to the observations, I
applied to the former a similar analysis, using the same
software. The full procedure is described in \inlinecite{Ath04a}. 
The results for two
different viewing angles are shown in Fig.~\ref{fig:unsharp} for a
simulation with a strong bar. The
similarity between the features in the median filtered 
observations and simulations is very striking.

In the upper panel of Fig.~\ref{fig:unsharp} the disc is viewed
edge-on, with the bar side-on. It displays a clear X-like form, and
 secondary maxima on the equatorial plane
on either side of the center, in good agreement with median filtered
images of observed galaxies with peanuts.
It further shows two very faint features, like
parentheses enclosing the X.
The lower panel of Fig.~\ref{fig:unsharp} 
shows the same simulation but from a viewing angle of  $45^{\circ}$.
Viewed side-on, the four branches of the X do not cross
the center. This is probably still true, but less easy to see, when the
viewing angle is $45^{\circ}$. Furthermore, the outermost isodensity
contours joining the two upper (or lower) branches of the X look
curved. This can also be seen in a number of the median filtered 
galaxy images in \inlinecite{Aron04}, which might mean that
these galaxies are seen from a viewing angle around
$45^{\circ}$. Finally, the secondary 
maxima along the equatorial plane can be seen from both viewing
angles (as well as from an end-on view not shown here), 
but are most prominent in the side-on view. All
these features were seen in the median filtered images of the galaxies
in \inlinecite{Aron04}. They are not accidental; they correspond to
specific structures of the periodic orbits that constitute the
backbone of barred galaxies \cite{PSA02}. A more
thorough comparison of the observations, the simulations and the
periodic orbit structure will be given elsewhere (see also 
\opencite{Aron04}).

\begin{figure} 
\rotatebox{0}{\includegraphics[width=25pc]{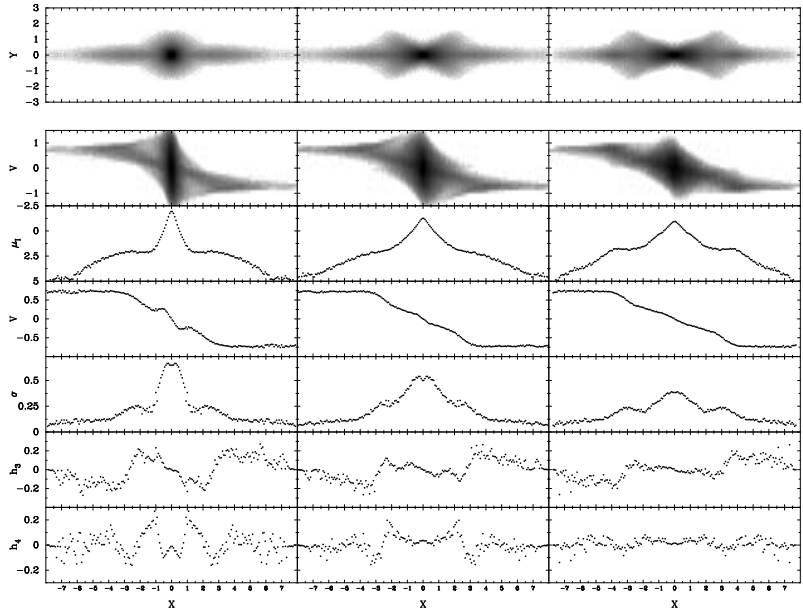}}
\caption[]{Analysis of the kinematic information obtained along
the equatorial plane of an edge-on $N$-body disc galaxy with a
strong bar. The three
columns correspond to different viewing angles. In all three the
disc is seen edge-on. In the left panels the bar is seen end-on, 
in the right panels side-on, and in the central ones the 
line-of-sight is oriented at 45$^{\circ}$ with respect to the bar
major axis. The upper row gives a greyscale representation of the
disc material. The second row gives the corresponding position - 
velocity diagram, again in greyscale. The third row gives the surface
brightness as function of distance along the equatorial plane. The
fourth, fifth, sixth, and seventh rows give the velocity, $\sigma$,
$h_3$ and $h_4$, respectively, where $\sigma$ is the velocity dispersion,
and $h_3$ and $h_4$ are the 3rd and 4th components of the Gauss-Hermite
series.}
\label{fig:pvd}
\end{figure}
 
\subsection{PVD diagrams }

Emission line spectroscopy of boxy/peanut galaxies (\citeauthor{KM95} 
\citeyear{KM95}, \citeauthor{MK99} \citeyear{MK99}, 
\citeauthor{BF99} \citeyear{BF99}) showed that their major 
axis position velocity diagrams (hereafter PVDs) show a number of interesting
features which were first linked to bars by \citeauthor{KM95}. 
\inlinecite{BA99} used periodic orbits in a  
standard barred galaxy potential as a  first step in understanding 
such features. Although this approach is too crude 
to reproduce, even roughly, observed PVDs, it can still give useful insight.  
For example. it showed that the superposition of the main periodic orbit 
families leads to gaps between the signatures of the different
orbit families, as well as to material in the so-called forbidden
quadrants. To actually model emission line PVDs, \inlinecite{AB99}
used the gas flow 
simulations of \inlinecite{Ath92} viewed edge-on. They found that 
shocks along the leading edges of the bar and the corresponding inflow
lead to a characteristic gap in the PVDs, between the signature of the
nuclear spiral (whenever existent) and the signature of the
disc. There is in general very good agreement between the signatures
of the observed emission line PVDs and those obtained by hydrodynamic
models. 

\inlinecite{CB04} presented long-slit absorption line kinematic
observations of 30 edge-on disc galaxies, most of which have a boxy 
structure or a peanut. They analysed the corresponding PVDs using
Gauss-Hermite series and produced profiles of the integrated
light along the slit, the mean stellar velocity $V$, the velocity dispersion
$\sigma$, as well as the higher order components $h_3$ and $h_4$. 
\inlinecite{BA04} `observed' 
many $N$-body bars seen edge-on, using   
for their analysis exactly the same techniques and, whenever possible,
also the same software. It is thus 
both straightforward and meaningful to make comparisons of the
results of the two studies.

The similarities are striking. Both studies find the same
characteristic signatures, one in the observations and the other in
the simulations. Fig.~\ref{fig:pvd} 
gives the results for an $N$-body simulation 
with a strong bar. The 
integrated light along the slit (equivalent to a major-axis light
profile) has a quasi-exponential central component and a plateau at
intermediate radii, followed by a steep drop. The rotation curve shows a
double hump. The velocity dispersion has a  
central peak, which in the center-most part may be rather flat or may
even have a central minimum. At intermediate radii there can be a plateau
which sometimes ends on either side with a shallow maximum before
dropping steeply at larger radii. In fact, these features can already
be seen in Figure 13 of \inlinecite{AM02}, although the display
adopted in that paper makes it more difficult to compare directly to
observations. Fig.~\ref{fig:pvd} also shows that  
$h_3$ correlates with $V$ over most of the bar
length, contrary to what is expected for a fast rotating disc. All
these features are spatially correlated and are seen, more 
or less strongly, both in the 
observations and in the simulations. 
The $N$-body simulations show clearly that the strength of these 
features depends on the strength of the bar as well as on its viewing
angle. Furthermore, those features
can be interpreted with the help of the orbital structure in barred
discs \cite{BA99}.

The only observed feature not found by \inlinecite{BA04}
is the anti-correlation of $h_3$ and $V$ in the
center-most parts. Indeed, the observations show that, in a small
region very near the
center, the $h_3$ and $V$ curves anti-correlate in many cases,
while for the 
remaining bar region they correlate. On the hand, in the
simulations shown by \inlinecite{BA04} the $h_3$ and $V$
curves always correlate all the way to the center. This small discrepancy can
be remedied in two ways, presented and discussed elsewhere 
(\citeauthor{Ath04b}, in preparation). 
The first relies on the existence of a gaseous inner
disc, present in the galaxies but not in the simulations discussed
by \inlinecite{BA04}, while the
second relies on a more centrally concentrated halo profile. Thus
even this small discrepancy between observations and simulations can
be remedied.

\section{Summary and discussion}

Bars are present in the majority of observed disc galaxies and also 
form spontaneously in $N$-body simulations. Contrary to previous beliefs, 
it is now well established that haloes can destabilise bars, since they can 
take positive angular momentum from them and thus make them stronger and 
slower rotating. This of course will work only if all components are 
non-rigid and capable of emitting/absorbing angular momentum.

It is possible to account for the considerable spread in observed bar 
morphologies. These can range from short bars or ovals, which should be found 
in disc galaxies within which not much angular momentum has been exchanged, 
to very strong bars with rectangular-like isophotes, which on the contrary 
should be in galaxies having undergone considerable amount of angular 
momentum redistribution. Extreme cases of such galaxies are some examples 
of bar dominated 
early type discs presented by \inlinecite{Gad03}, in which the 
disc is not a major component any more, since 
a very considerable fraction of its mass is now within the bar component.
Indeed, it is possible to turn the problem around and to try and obtain 
some information on galaxy haloes from the morphology of their bars. Even
though no strong conclusions can be reached in this way, the insight thus
obtained is still very welcome, since our knowledge on disc galaxy haloes
is still very restricted. Thus short bars, like in our own Galaxy, should 
be indicative of very hot haloes, which could not absorb much angular 
momentum. On the contrary, strong and long bars, should be indicative 
of haloes that can absorb angular momentum, with considerable amount 
of material in the relatively inner regions where the resonances lie. 

A short while after having formed, the bar buckles perpendicular to the 
disc equatorial plane and forms an edge-on boxy or peanut shape. Detailed 
comparison of $N$-body bars with structures known as `boxy or peanut bulges' 
-- including morphology, photometry and kinematics --
show that these so-called bulges are in fact the part of the bar 
sticking out of the equatorial plane of the galaxy. This behaviour and 
structure is easily understood with the help of the orbital structure in 
3D barred galaxy potentials.

\appendix

In this appendix I summarise some elementary notions for orbits and
resonances in order to introduce the notation and nomenclature used
in this paper. 

\begin{figure} 
\rotatebox{0}{\includegraphics[width=27pc]{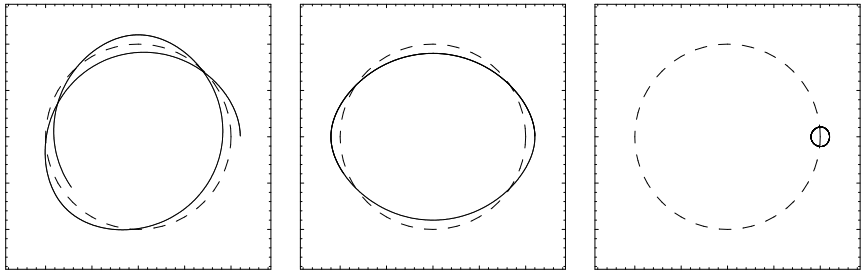}}
\caption[]{Three examples of epicyclic orbits. In the middle panel 
we have $l = -1$ and $m = 2$, so that the orbit is at the ILR. In the
right panel we have $l = 0$, so that the orbit is at corotation.
In the left panel the $l$ and $m$ values are not commensurable, so that
the orbit does not close.}
\label{fig:epicorb}
\end{figure}

In galactic discs, many stars are on orbits which are not too
far from circular, so that the epicyclic approximation is valid.
Orbits on the equatorial plane of a
disc galaxy are then characterised by two frequencies : the angular
frequency $\Omega$ and the frequency of the radial motion, called
epicyclic frequency, $\kappa$. In a frame
of reference in which the bar (or, more generally, the
non-axisymmetric perturbation) is at rest, the angular frequency
becomes $\Omega - \Omega_p$, where $\Omega_p$ is the pattern
speed. Resonances occur when the angular and radial 
frequencies are commensurable, i.e. when 

\begin{equation}
\frac{\Omega - \Omega_p}{\kappa} = - \frac{l}{m},
\end{equation}

\noindent
where $l$ and $m$ are integers,
so that the orbit closes in the rotating frame of reference. If the
orbit closes after one rotation and two radial oscillations, i.e. when
$l = - 1$ and $m$ = 2, it is at inner Lindblad resonance
(ILR). An example of such an orbit is shown in the middle panel of
Fig.~\ref{fig:epicorb}. For $l$ = 1 and $m$ = 2 the orbit is at the outer
Lindblad resonance (OLR). If the orbit is at corotation (CR), then $l$
= 0 (right panel of Fig.~\ref{fig:epicorb}). Further resonances of
importance to bars are the inner ultra-harmonic resonance (iUHR) with
$l = -1$ and $m = 4$, and in general all resonances with $l = -1$.
If the $l$ and $m$ values
are not commensurable the orbit will not close (left panel of
Fig.~\ref{fig:epicorb}). Epicyclic orbits in three dimensions 
have one more frequency, the frequency of the vertical oscillation
($\kappa_z$), so that further resonances are possible. For example if   

\begin{equation}
\frac{\Omega - \Omega_p}{\kappa_z} = - \frac{n}{m}
\end{equation}

\noindent
with $n = -1$ and $m$ = 2, then the orbit is at the vertical ILR,
often named the 2:1 vertical resonance. Vertical resonances of the m:1 type
are in general important since they can introduce families which are
crucial for building the peanut structure.

Of course, in general orbits are not simple epicycles, yet working
in action-angle variables (e.g. \citeauthor{BinTrem}, \citeyear{BinTrem}) 
makes it possible
to deal with the regular orbits in ways similar to those used for
epicycles, and in particular to define resonances in a similar way.
It is also possible to determine the principal frequencies of an
orbit numerically \cite{BinSper} and this can be made even for
potentials which are not known in a closed analytical form, as e.g.
those from $N$-body simulations.

\acknowledgements
In the past four years, during which I was working on the
subject matters discussed in this paper, I benefitted from a number of
interesting 
and motivating discussions with A. Bosma, M. Bureau, W. Dehnen,
A. Misiriotis, P. Patsis, M. Tagger, 
G. Aronica, I. Berentzen, K. Freeman, B. Fuchs,
A. Kalnajs, J. Kormendy, D. Lynden-Bell, F. Masset,
and Ch. Skokos. I thank Jean-Charles
Lambert for his invaluable help with the simulation software and the
administration of the runs and W. Dehnen for 
making available to me his tree code and related programs. 
I also thank the Observatoire de Marseille, the region PACA, the
INSU/CNRS and the University of Aix-Marseille I for funds to develop
the computing facilities used for the calculations in this paper.

\end{article}
\end{document}